\documentclass[3p,times]{elsarticle}

\usepackage{ecrc}

\volume{00}

\firstpage{1}

\journalname{Solid State Communications}

\runauth{}

\jid{procs}

\jnltitlelogo{Solid State Commun.}

\CopyrightLine{2014}{Published by Elsevier Ltd.}

\usepackage{amssymb}

\usepackage[figuresright]{rotating}
\usepackage[english]{babel}

\begin{document}

\begin{frontmatter}




\title{The Kohn-Luttinger superconductivity in idealized doped graphene}

\author[label1,label2]{M. Yu. Kagan}
\ead{kagan@kapitza.ras.ru}
\author[label3]{V. V. Val'kov}
\ead{vvv@iph.krasn.ru}
\author[label3,label4]{V. A. Mitskan}
\author[label3]{M. M. Korovushkin}
\address[label1]{P.\,L. Kapitza Institute for Physical Problems, 119334 Moscow, Russia}
\address[label2]{Moscow Institute of Electronics and Mathematics,
National Research University Higher School of Economics, 109028
Moscow, Russia}
\address[label3]{L.\,V. Kirensky Institute of Physics, 660036 Krasnoyarsk, Russia}
\address[label4]{Siberian State Aerospace University, 660014 Krasnoyarsk,
Russia}

\begin{abstract}
Idealized graphene monolayer is considered neglecting the van der
Waals potential of the substrate and the role of the nonmagnetic
impurities. The effect of the long-range Coulomb repulsion in an
ensemble of Dirac fermions on the formation of the superconducting
pairing in a monolayer is studied in the framework of the
Kohn-Luttinger mechanism. The electronic structure of graphene is
described in the strong coupling Wannier representation on the
hexagonal lattice. We use the Shubin-Vonsowsky model which takes
into account the intra- and intersite Coulomb repulsions of
electrons. The Cooper instability is established by solving the
Bethe-Salpeter integral equation, in which the role of the
effective interaction is played by the renormalized scattering
amplitude. The renormalized amplitude contains the Kohn-Luttinger
polarization contributions up to and including the second-order
terms in the Coulomb repulsion. We construct the superconductive
phase diagram for the idealized graphene monolayer and show that
the Kohn-Luttinger renormalizations and the intersite Coulomb
repulsion  significantly affect the interplay between the
superconducting phases with $f-$, $d+id-$, and $p+ip-$wave
symmetries of the order parameter.

\end{abstract}


\end{frontmatter}


\section{Introduction}
\label{intro}

One of the most interesting properties of graphene is
controllability of the position of its chemical potential by an
applied electric field, which allows the change of the carrier
type (electrons or holes)~\cite{Lozovik08,Castro09}. It was
experimentally demonstrated that short graphene samples placed
between superconducting contacts could be used for constructing
Josephson junctions~\cite{Heersche07}. This indicates that Cooper
pairs can coherently propagate in graphene. The question now
arises of whether graphene can be structurally or chemically
modified to become a magnet~\cite{Peres05} or even a true
superconductor.

Theoretically, a model with the conical dispersion requires the
minimum intensity of the pairing interaction to develop the Cooper
instability~\cite{Marino06}. In view of this fact, a number of
attempts were made to theoretically analyze possible
implementation of the superconducting state in doped graphene. In
paper~\cite{Gonzalez01}, the role of topological effects in
implementation of the Cooper pairing in this material was
investigated. In paper~\cite{Uchoa07}, using the mean field
approximation, the plasmon type of superconductivity in graphene
was investigated, which leads to the low critical temperatures in
the $s-$wave channel for realistic electron densities. The
possibility of inducing superconductivity in graphene by electron
correlations was studied in~\cite{Black07,Honerkamp08}. In
paper~\cite{Kiesel12}, the interplay of the superconducting phase
with the $d+id-$wave symmetry of the order parameter and the spin
density wave phase depending on the position of  the chemical
potential with respect to van Hove singularity in the electron
density of states of graphene was investigated using the
functional renormalization group. Near the van Hove singularity,
the superconducting phases with $d+id-$ and $f-$wave symmetries of
the order parameter were found.

In paper~\cite{Gonzalez08}, the situation was considered when the
Fermi level is located near one of the van Hove singularities in
the density of states of graphene. It is known that these
singularities can enhance the magnetic and superconducting
fluctuations~\cite{Markiewicz97}. According to the scenario
described in~\cite{Gonzalez08}, the Cooper instability occurs due
to the strong anisotropy of the Fermi contour at van Hove filling
$n_{vH}$, which, as a matter of fact, originates from the
Kohn-Luttinger mechanism~\cite{Kohn65} proposed in 1965 and
suggesting the appearance of the superconducting pairing in
systems with the purely repulsive interaction. According to the
estimation made in~\cite{Gonzalez08}, the Cooper instability of
this type in idealized graphene can increase the critical
temperatures of the superconducting transition up to 10 K,
depending on whether the chemical potential level is close to the
van Hove singularity. It should be noted that in the calculation
only the Coulomb repulsion of electrons on one site was taken into
account. In paper~\cite{Valenzuela08}, the possible interplay and
coexistence of the Pomeranchuk instability and the Kohn-Luttinger
superconducting pairing in graphene were discussed. The authors
of~\cite{Nandkishore12} demonstrated using a renormalization group
approach within the Kohn-Luttinger mechanism that in a monolayer
of the doped graphene  the superconducting $d+id-$pairing can be
implemented.

In this paper, an idealized monolayer of graphene is considered
neglecting the van der Waals potential of the substrate and the
role of the nonmagnetic impurities. The Cooper instability in a
monolayer is investigated in the weak coupling limit of the Born
approximation by implementing the Kohn-Luttinger mechanism with
respect to the Coulomb repulsion of electrons localized not only
on one, but also on the nearest-neighbor carbon atoms. In the
evaluation of the effective interaction in the Cooper channel, we
take into account the polarization contributions caused by the
Coulomb repulsion between electrons belonging to both one and
different branches of the graphene energy spectrum.

The necessity to account for the long-range Coulomb repulsion in
the calculation of the physical characteristics of graphene was
dictated by the results of paper~\cite{Wehling11}, where in the ab
initio calculation of the effective many-body model of graphene
and graphite the values of the partially screened
frequency-dependent Coulomb repulsion were determined. It was
demonstrated that the value of the onsite repulsion in graphene is
$U=9.3$ eV and the Coulomb repulsion of electrons localized on the
neighboring sites is $V=5.5$ eV, which indicates the principle
importance to take into account the nonlocal Coulomb interaction.
Note that other researches consider the values of $U$ and $V$ to
be much smaller.

\section{Theoretical model}
\label{model}

Since there are two carbon atoms per each unit cell of the
graphene lattice, the latter can be divided in two sublattices A
and B. In the Wannier representation, the Hamiltonian of the
Shubin-Vonsowsky model (the extended Hubbard
model)~\cite{Shubin34} for graphene with respect  to electron
hoppings between the nearest-neighbor and next to-nearest-neighbor
atoms and the Coulomb repulsion of electrons located at one and at
neighboring sites has the form
\begin{eqnarray}
\hat{H}&=&\hat{H}_0+\hat{H}_{int},\\
\hat{H}_0&=&-\mu\sum_{f} (\hat{n}^{A}_{f}+
\hat{n}^{B}_{f})-t_1\sum_{\langle
fm\rangle\sigma}(a^{\dag}_{f\sigma}b_{m\sigma}+\textrm{h.c.})\nonumber\\
&&-t_2\sum_{\langle\langle
fm\rangle\rangle\sigma}(a^{\dag}_{f\sigma}a_{m\sigma}+
b^{\dag}_{f,\sigma}b_{m,\sigma}+\textrm{h.c.}),\\\label{Hint}
\hat{H}_{int}&=&U\sum_{f}
(\hat{n}^{A}_{f\uparrow}\hat{n}^{A}_{f\downarrow}+
\hat{n}^{B}_{f\uparrow}\hat{n}^{B}_{f\downarrow})+V\sum_{\langle
fm\rangle} \hat{n}^{A}_{f}\hat{n}^{B}_{m}.
\end{eqnarray}
Here, $a^{\dag}_{f\sigma}(a_{f\sigma})$ are the operators that
create (annihilate) an electron with the spin projection
$\sigma=\pm1/2$ at site $f$ of the sublattice A,
$\displaystyle\hat{n}^{A}_{f}=\sum_{\sigma}\hat{n}^{A}_{f\sigma}=\sum_{\sigma}a^{\dag}_{f\sigma}a_{f\sigma}$
are the operators of the numbers of fermions at site $f$ of the
sublattice A (the analogous notations are used for the sublattice
B), $\mu$ is the chemical potential of the system, $t_1$ is the
hopping integral  between neighboring atoms (hoppings between
different sublattices), $t_2$ is the hopping integral between the
next to nearest-neighbor atoms (within one sublattice), $U$ is the
parameter of the Coulomb repulsion of electrons located at one
site and having the opposite spin projections (Hubbard repulsion),
and $V$ is the Coulomb repulsion of electrons located at
neighboring atoms. In the Hamiltonian, $\langle~\rangle$ denotes
the summation over the nearest neighbors only,
$\langle\langle~\rangle\rangle$ -- the summation over the next to
nearest neighbors.

After the transition to the momentum state and the Bogoliubov
transformation
\begin{eqnarray}\label{uv}
\alpha _{i,k,\sigma}= w_{i1}(k){a_{ k,\sigma }} +
w_{i2}(k){b_{k,\sigma }},\qquad i=1,2,
\end{eqnarray}
the Hamiltonian $\hat{H}_0$ is diagonalized and acquires the form
\begin{eqnarray}
\hat H_0 =\sum\limits_{i=1}^2 \sum\limits_{ k\sigma } E_{i,k}
{\alpha_{i,k,\sigma}^{\dag}\alpha_{i,k,\sigma}}.
\end{eqnarray}
The two-band energy spectrum of graphene is described by the expressions~\cite{Wallace47}
\begin{eqnarray}\label{spectra}
E_{1,k}=t_1|u_k|-t_2f_k,\qquad E_{2,k}=-t_1|u_k|-t_2f_k,
\end{eqnarray}
where the notations
\begin{eqnarray}\label{f_k}
&&f_k=2\cos(\sqrt{3}k_y)+
4\cos\biggl(\frac{\sqrt{3}}{2}k_y\biggr)\cos\biggl(\frac{3}{2}k_x\biggr),\\
&&u_k=\displaystyle\sum_{\delta}e^{i k\delta}=e^{-ik_x}+
2e^{\frac{i}{2}k_x}\cos(\frac{\sqrt{3}}{2}k_y),\qquad
|u_k|=\sqrt{3+f_k}\nonumber
\end{eqnarray}
were used. The Bogoliubov transformation parameters have the form
\begin{eqnarray}\label{wz}
&&{w_{1,1}}(k) = w_{22}^*(k) = \frac{1}{\sqrt{2}}r_k^*,\qquad r_k=\frac{u_k}{|u_k|},\\
&&{w_{12}}(k) = -w_{21}(k) =-\frac{1}{\sqrt{2}}.\nonumber
\end{eqnarray}

In the Bogoliubov representation of quasiparticles, the
interaction operator (\ref{Hint}) is determined by the expression
containing $\alpha_{1,k,\sigma}$ and $\alpha_{2,k,\sigma}$
\begin{eqnarray}\label{Hint_ab}
\hat H_{int} &=& \frac{1}{N}\sum\limits_{i,j,l,m\atop
k,p,q,s,\sigma} \Gamma_{ij;lm}^{||}(kp|qs)
\alpha_{ik\sigma}^\dag \alpha_{jp\sigma}^\dag \alpha_{lq\sigma}\alpha_{ms\sigma}\Delta (k+p-q-s) +\\
 &+& \frac{1}{N}\sum\limits_{i,j,l,m\atop k,p,q,s}\Gamma_{ij;lm}^{\bot}(kp|qs)
\alpha_{ik\uparrow}^\dag \alpha_{jp\downarrow}^\dag \alpha_{lq\downarrow}\alpha_{ms\uparrow}\Delta (k+p-q-s),\nonumber
\end{eqnarray}
where the initial amplitudes
\begin{equation}
\Gamma_{ij;lm}^{||}(kp|qs) = V_{ij;lm}(kp|qs) =
V {u_{q-p}} w_{i1}(k) w_{j2}(p) w^*_{l2}(q) w^*_{m1}(s),
\end{equation}
describe the intensity of the interaction of Fermi quasiparticles
with the parallel spins and the initial amplitudes
\begin{eqnarray}
&&\Gamma_{ij;lm}^{\bot}(kp|qs) = V_{ij;lm}(kp|qs) + V_{ji;ml}(pk|sq) + U_{ij;lm}(kp|qs);\\
&&\qquad U_{ij;lm}(kp|qs) =
U\Bigl( w_{i1}(k) w_{j1}(p) w^*_{l1}(q) w^*_{m1}(s) +
w_{i2}(k) w_{j2}(p) w^*_{l2}(q) w^*_{m2}(s) \Bigr),\nonumber
\end{eqnarray}
describe the interaction of Fermi quasiparticles with antiparallel
spins. Indices ${i,j,l,m}$ can take the values of 1 or 2. Note
that as far as the terms $\alpha_{ik\sigma}^\dag
\alpha_{jp\sigma}^\dag \alpha_{lq\sigma}\alpha_{ms\sigma}$ and
$\alpha_{jp\sigma}^\dag\alpha_{ik\sigma}^\dag
\alpha_{ms\sigma}\alpha_{lq\sigma}$ correspond to the same
process, the effective interaction $\Gamma^{||}$ should be written
as
\begin{equation}
\Gamma_{ij;lm}^{||}(kp|qs) = V_{ij;lm}(kp|qs) +
(1-\delta_{ij}\delta_{lm})V_{ji;ml}(pk|sq).
\end{equation}

\section{Effective interaction in the Cooper channel
and the equation for the order parameter }
\label{renorminteraction}

\begin{figure}
\begin{center}
\includegraphics[width=0.55\textwidth]{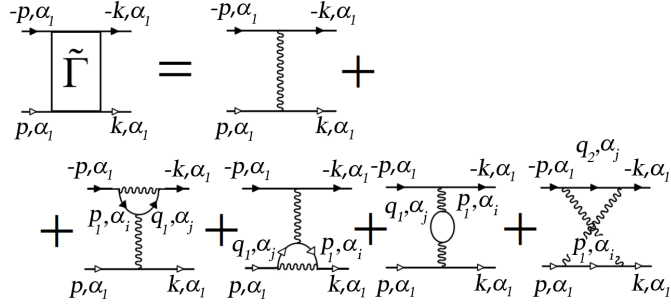}
\caption{\label{diagrams_alpha}First- and second-order diagrams
for the effective interaction of electrons. Solid lines with the
light (dark) arrows correspond to the Green's functions of
electrons with spin projections equal to $+{\textstyle{1 \over
2}}$~($-{\textstyle{1 \over 2}}$) and the energy corresponding to
the upper ($\alpha_1$) or lower ($\alpha_2$) bands in graphene.
Indices $i$ and $j$ acquire the values 1 or 2. The momenta $q_i$
are defined in Eq. (\ref{q1q2}).}
\end{center}
\end{figure}

The utilization of the weak coupling Born approximation in the
evaluation  of the scattering amplitude in the Cooper channel
allows us to limit the consideration up to the second order
diagrams in  the effective interaction for two electrons with the
opposite values of the momentum and spin and use the quantity
$\widetilde{\Gamma}( p, k)$. This quantity is graphically
determined as a sum of the diagrams shown in
Fig.~\ref{diagrams_alpha}. Solid lines with the light (dark)
arrows correspond to the Green's function of the electrons with
spin projections equal to $+\frac12 ~(-\frac12)$. It is well-known
that the possibility of the Cooper pairing is determined by the
characteristics of the energy spectrum close to the Fermi level
and the effective interaction of electrons located near the Fermi
surface ~\cite{Gor'kov61}. Assuming that upon doping of graphene
the chemical potential moves in the upper energy band $E_{1,k}$
and analyzing the conditions for the appearance of the
Kohn-Luttinger superconductivity we may consider that the initial
and final momenta will also belong to the upper band. This is
shown in Fig.~\ref{diagrams_alpha} by indices $\alpha_1$ (upper
band) and $\alpha_2$ (lower band).

The first plot in Fig.~\ref{diagrams_alpha} corresponds to the
bare interaction of two electrons in the Cooper channel and is
determined analytically by the expression
\begin{eqnarray}
\widetilde{\Gamma}_0(p,k)=\frac U2+
\frac V4\Bigl(u_{p-k} r^*_p r_k+\textrm{h.c.}\Bigr),
\end{eqnarray}
where we took into account that $u_{-k}=u^*_k$. The next
(Kohn-Luttinger) diagrams in Fig. 1 originate from the
second-order scattering processes, $\delta\widetilde{\Gamma}( p,
k)$, and take into account the polarization effects of the filled
Fermi sphere. In these diagrams, the presence of the two solid
lines without arrows indicates the performed summation over the
both values of the spin projections. Wavy lines correspond to the
bare interaction. The scattering of electrons with the same spin
projection gives rise only to the intersite contribution. If we
have the interaction between electrons with opposite spins, the
scattering amplitude is determined by the sum of the Hubbard and
intersite repulsions. Therefore, when we deal only with the
Hubbard repulsion, the $\delta\widetilde{\Gamma}( p, k)$
correction for the effective interaction is given only by the last
diagram of the exchange type. If we take into account the Coulomb
repulsion at the neighboring sites, then all the diagrams in
Fig.~\ref{diagrams_alpha} contribute to the renormalized
amplitude.

After the introduction of the  analytical expression for the
diagrams, we perform the summation over the Matsubara frequencies.
Here, we take into account that the main contribution to the total
scattering amplitude $\Gamma(p\,|k)$ in the Cooper channel comes
from the scattering of electrons with the energies close to the
Fermi energy, therefore, we can ignore the Matsubara frequency
dependence of $\widetilde{\Gamma}$ in the Bethe-Salpeter integral
equation. As a result, we get the following integral expression
for the effective interaction
\begin{eqnarray}\label{Gamma_wave}
\widetilde{\Gamma}(p,k)&=&
\widetilde{\Gamma}_0(p,k)+\delta\widetilde{\Gamma}(p,k).
\end{eqnarray}
The total contribution of the second-order diagram yields
\begin{eqnarray}\label{deltaGamma}
\delta \tilde \Gamma (p,k)=
- \frac{1}{N}\sum\limits_{i,j, p_1} \chi_{i,j}(q_2, p_1)
\Gamma^{\bot}_{1i;1j}(p, q_2| -k, p_1)
\Gamma^{\bot}_{j1;i1}(p_1, -p|q_2,k)\nonumber \\
- \frac{1}{N}\sum\limits_{i,j, p_1} \chi_{i,j}(q_1,p_1) \Bigl\{
\Gamma^{\bot}_{1j;i1}( p, p_1| q_1, k)\left[\Gamma^{||}_{i1;j1}(
q_1, -p| p_1, -k) -
\Gamma^{||}_{i1;1j}( q_1, -p| -k, p_1) \right] \Bigr.\\
\qquad\qquad\Bigl. + \Gamma^{\bot}_{i1;1j}(q_1, -p| -k, p_1) \left[\Gamma^{||}_{1j;1i}( p, p_1|  k, q_1) -
\Gamma^{||}_{1j;i1}( p, p_1|  q_1, k)\right]  \Bigr\}.\nonumber
\end{eqnarray}
Here, the notations for the generalized susceptibilities
\begin{equation}
\chi_{i,j}(k,p) =
\frac{f(E_{i,k}) - f(E_{j,p})}
{E_{i, k} - E_{j,p}},
\end{equation}
are used, where $f(x)=(\exp(\frac{x-\mu}{T})+1)^{-1}$ is the
Fermi-Dirac distribution  function and the energies $E_{i,k}$ are
given by formulas~(\ref{spectra}). For the sake of compactness, we
introduce the following notations of the momenta combinations
\begin{eqnarray}\label{q1q2}
q_1 =  p_1 + p - k;\qquad q_2 = p_1-p-k.
\end{eqnarray}

Knowing the renormalized expression for the effective interaction,
we may proceed to the analysis of the conditions for the
realization of the Cooper instability in the investigated model.
It is known~\cite{Gor'kov61} that the appearance of the Cooper
instability can be found by analyzing the homogeneous part of the
Bethe-Salpeter equation. In this case, the dependence of the
scattering amplitude $\Gamma(p,k)$ on momentum  $k$ is factorized
and the integral equation for the superconducting gap $\Delta(p)$
is obtained. Introducing the integration over the isoenergetic
contours, we reduce the investigation of the Copper instability to
the solution of an eigenvalue problem
~\cite{Scalapino86,Baranov92,Hlubina99,Raghu10,Kagan13a,Kagan13b}
\begin{equation}
\label{IntegralEqPhi}
\frac{1}{(2\pi)^2}\oint\limits_{\varepsilon_{\vec{q}}=\mu}
\frac{d\hat{q}} {v_F(\hat{q})}
\widetilde{\Gamma}(\hat{p},\hat{q})
\Delta(\hat{q})=\lambda\Delta(\hat{p}),
\end{equation}
where the superconducting order parameter $\Delta(\hat{q})$ plays
the role of the eigenvector and the eigenvalues $\lambda$ satisfy
the relation $\lambda^{-1}\simeq \ln(T_c/W)$. Here $W$ is a
bandwidth both for the upper and the lower branches of graphene
energy spectrum determined by Eqs.~(\ref{spectra})--(\ref{f_k}) in
case when $t_2 =0$. In this case, the momenta $\hat{p}$ and
$\hat{q}$ lie on the Fermi surface and $v_F(\hat{q})$ is the Fermi
velocity.

To solve Eq.~(\ref{IntegralEqPhi}), we represent its kernel as a
superposition of the eigenfunctions each belonging to one of the
irreducible representations of the $C_{6v}$ symmetry group on the
hexagonal lattice. It is known that this group has six irreducible
representations ~\cite{Landau77}: four one-dimensional and two
two-dimensional. For each representation,
Eq.~(\ref{IntegralEqPhi}) has a solution with its own effective
coupling constant $\lambda$. Further on, we use the following
notation to classify the symmetries of the order parameter:
representation $A_1$ corresponds to the $s-$wave symmetry; $B_1$
and $B_2$, to the $f-$wave symmetry; $E_1$, to the $p+ip-$wave
symmetry, and $E_2$, to the $d+id-$wave symmetry.

For the irreducible representation $\nu$, we search a solution of
Eq.~(\ref{IntegralEqPhi}) in the form
\begin{equation}\label{solution}
\Delta^{(\nu)}(\phi)=\sum\limits_{m}\Delta_{m}^{(\nu)}g_{m}^{(\nu)}(\phi),
\end{equation}
where $m$ is the number of an eigenfunction belonging to the
representation $\nu$ and $\phi$ is the angle defining the
direction of the momentum $\hat{p}$ with respect to the $p_x$
axis. The explicit form of the orthonormalized functions
$g_{m}^{(\nu)}(\phi)$ is determined by the expressions
\begin{eqnarray}\label{harmon}
&&A_1\rightarrow~g_{m}^{(s)}(\phi)=\frac{1}{\sqrt{(1+\delta_{m0})\pi}}\,
\textrm{cos}\,6m\phi,~~m\in[\,0,\infty),\label{invariants_s}\nonumber\\
&&A_2\rightarrow~g_{m}^{(A_2)}(\phi)=\frac{1}{\sqrt{\pi}}\,\textrm{sin}\,
(6m+6)\phi,\label{invariants_s1}\nonumber\\
&&B_1\rightarrow~g_{m}^{(f_1)}(\phi)=\frac{1}{\sqrt{\pi}}\,
\textrm{sin}\,(6m+3)\phi,\label{invariants_dxy}\\
&&B_2\rightarrow~g_{m}^{(f_2)}(\phi)=\frac{1}{\sqrt{\pi}}\,
\textrm{cos}\,(6m+3)\phi,\label{invariants_dx2y2}\nonumber\\
&&E_1~\rightarrow~g_{m}^{(p+ip)}(\phi)=\frac{1}{\sqrt{\pi}}\,(A\,\textrm{sin}\,
(2m+1)\phi+B\,\textrm{cos}\,(2m+1)\phi),\label{invariants_p}\nonumber\\
&&E_2~\rightarrow~g_{m}^{(d+id)}(\phi)=\frac{1}{\sqrt{\pi}}\,(A\,\textrm{sin}\,
(2m+2)\phi+B\,\textrm{cos}\,(2m+2)\phi)\label{invariants_p}\nonumber.
\end{eqnarray}
Here, for the two-dimensional representations $E_1$ and $E_2$,
index $m$ runs over the values at which the coefficients $(2m +
1)$ and $(2m+2)$, respectively, are not multiple of 3.

The eigenfunctions $g_{m}$ satisfy the orthonormality conditions
\begin{equation}\label{norma}
\int\limits_0^{2\pi}d\phi\,g_{m}^{(\nu)}(\phi)g_{
n}^{(\beta)}(\phi)=\delta_{\nu\beta}\delta_{mn}.
\end{equation}
\begin{figure}[h]
\begin{center}
\includegraphics[width=0.35\textwidth]{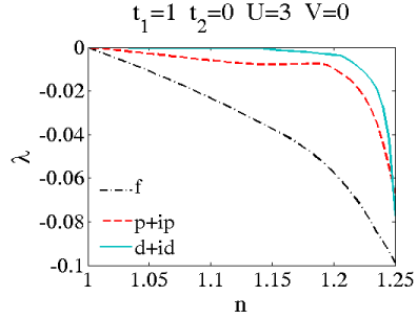}
\caption{\label{lambdas_aa}Dependences of $\lambda$ on the
electron density $n$ with respect  to the effective interaction of
the electrons with the energies corresponding to the upper branch
of the graphene energy spectrum $E_{1,k}$ for $t_2=0$, $U=3|t_1|$,
and $V=0$. The leading superconducting (SC) instability for all
the densities $1<n<1.25$ corresponds to $f-$wave pairing
($B_1-$representation of the order parameter, see
Eq.(\ref{harmon})).}
\end{center}
\end{figure}
\begin{figure}[h]
\begin{center}
\includegraphics[width=0.37\textwidth]{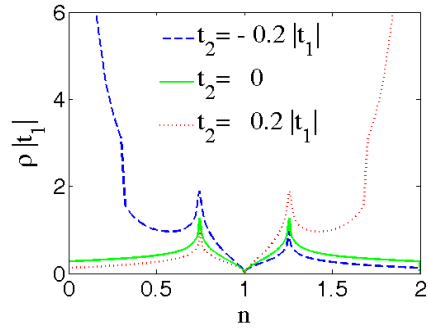}
\caption{\label{DOS} Evolution of the electron density of states
of graphene with the inclusion of hoppings to the next-to-nearest
neighbor atoms.}
\end{center}
\end{figure}

Substituting Eq.~(\ref{solution}) into Eq.~(\ref{IntegralEqPhi}),
performing integration over the angles, and using condition
(\ref{norma}), we find
\begin{equation}\label{EqDelta}
\sum_n\Lambda^{(\nu)}_{mn}\Delta_{n}^{(\nu)}=\lambda_{\nu}\Delta_{m}^{(\nu)},
\end{equation}
where
\begin{eqnarray}
\label{matrix}
\Lambda^{(\nu)}_{mn}&=&\frac{1}{(2\pi)^2}\oint\limits_0^{2\pi}d\phi_{\hat
p} \oint\limits_0^{2\pi}d\phi_{\hat q}\frac{d\hat{q}} {d\phi_{\hat
q}v_F(\hat{q})}
\widetilde{\Gamma}({\hat p},{\hat q})\nonumber\\
&\times&g_{m}^{(\nu)}(\phi_{\hat p}) g_{ n}^{(\nu)}(\phi_{\hat
q}).
\end{eqnarray}
Since $T_c\sim W\exp\bigl(1/\lambda \bigr)$, each negative
eigenvalue $\lambda_{\nu}$ corresponds to a superconducting phase
with the symmetry of the order parameter of the type $\nu$. The
expansion of the order parameter $\Delta^{(\nu)}(\phi)$ in terms
of the eigenfunctions generally includes many harmonics, but the
main contribution is made by several terms only. The highest
critical temperature corresponds to the largest absolute value of
$\lambda_{\nu}$.

\section{Results and Discussion}
\label{Results}

If upon doping of graphene the chemical potential moves to the
upper band $E_{1,k}$, then when we analyze the conditions for the
appearance of the Kohn-Luttinger superconductivity, we should
consider mostly the contribution of the scattering of the
electrons with the energies corresponding to the upper branch of
the energy spectrum (Eq.~(\ref{Gamma_wave}) at $i=j=1$).
Calculated dependences of the effective coupling constant
$\lambda$ on the carrier density $n$ for the different types of
symmetry of the superconducting order parameter are presented in
Fig.~\ref{lambdas_aa}. The calculation was performed for the set
of parameters $t_2=0,~U=3|t_1|$, and $V=0$. It can be seen that
over the entire region of the carrier density region $1<n<1.25$
the superconducting phase with the $f-$wave symmetry of the order
parameter is realized (the contribution comes from the
$g_{m}^{(f_1)}(\phi)=\displaystyle\frac{1}{\sqrt{\pi}}\,
\textrm{sin}\,(6m+3)\phi$ harmonics, whereas the contribution of
the $g_{m}^{(f_2)}(\phi)=\displaystyle\frac{1}{\sqrt{\pi}}\,
\textrm{cos}\,(6m+3)\phi$ is absent). Here and below, the figures
show only the curves corresponding to the $f-$, $p+ip-$,
$d+id-$symmetries which are characterized by the largest absolute
values of $\lambda_{\nu}$.
\begin{figure}[h]
\begin{center}
\includegraphics[width=0.35\textwidth]{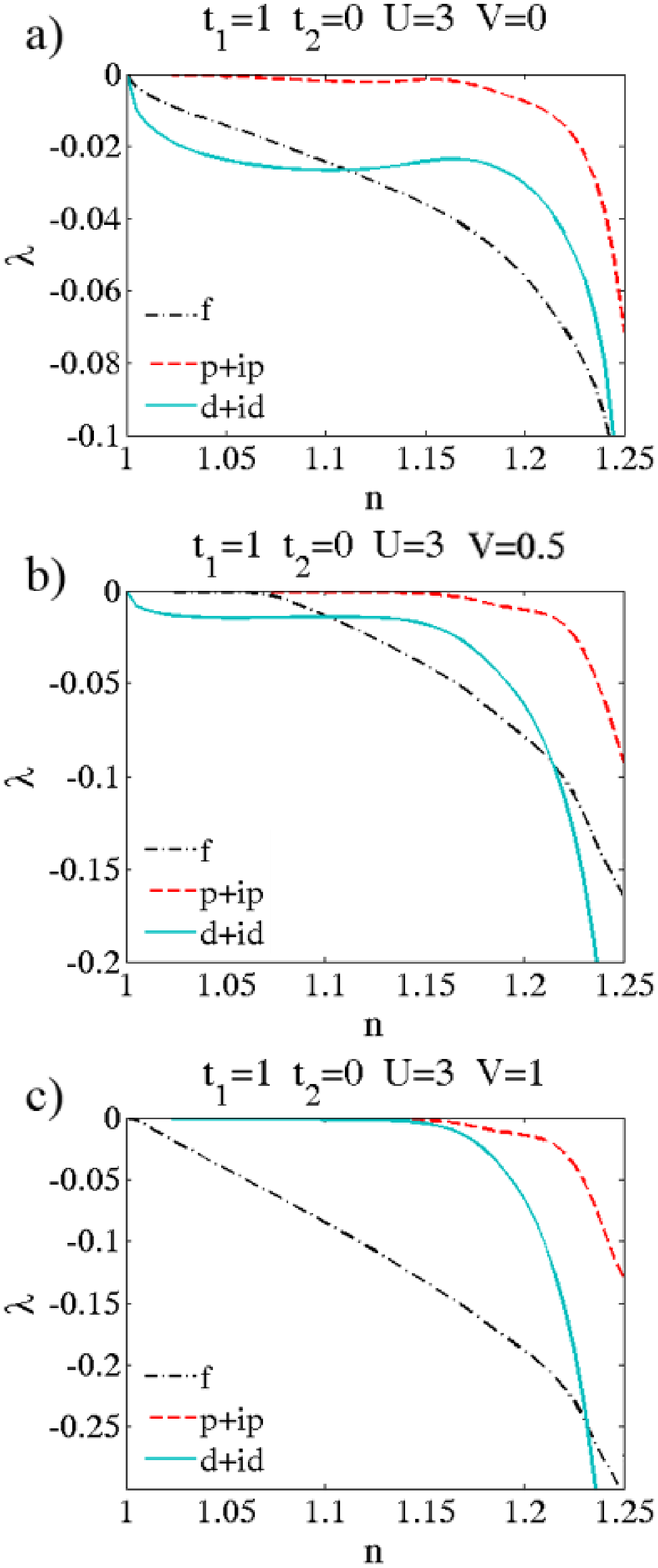}
\caption{\label{lambdas_ab}Dependences of $\lambda$ on the
electron density $n$ with respect to the effective interaction of
the electrons with the energies corresponding to both branches of
the graphene energy spectrum for $t_2=0$ and $U=3|t_1|$ at
different parameters of the intersite Coulomb repulsion: (a)
$V=0$, (b) $V=0.5|t_1|$, and (c) $V=1|t_1|$.}
\end{center}
\end{figure}

Note that in this paper, we analyze only the range of electron
densities for which we do not approach too close to the van Hove
singularity (solid green curve in Fig.~\ref{DOS}), in order to
escape the summation of parquet
diagrams~\cite{Dzyaloshinskii88a,Dzyaloshinskii88b}.

The inclusion of the Coulomb interaction of electrons with the
energies corresponding to different branches of the graphene
energy spectrum (in this case, the effective interaction is
described by the complete expression (\ref{Gamma_wave}))
qualitatively changes the superconducting phase diagram. In
particular, at the low electron densities $1<n<1.13$ and near the
van Hove singularity, the competition between the superconducting
phase with the $f-$wave and $d+id-$wave symmetries occurs, which
is described by the two-dimensional representation $E_2$
(Fig.~\ref{lambdas_ab}(a)). Namely for the densities $1<n<1.13$
the leading SC-instability corresponds to $d+id-$wave pairing,
while for the densities $1.13<n<1.25$ we have $f-$wave pairing. It
agrees well with the dependences $\lambda(n)$ calculated for the
Hubbard model on the hexagonal lattice in paper~\cite{Raghu10}.

The inclusion of the intersite Coulomb repulsion significantly
affects the interplay between the superconducting phases. This is
clearly seen in Fig.~\ref{lambdas_ab}(b) and (c), where we show
the dependences of $\lambda(n)$ for $V=0.5|t_1|$ and $V=1|t_1|$.
Their comparison to the plots in Fig.~\ref{lambdas_aa}
demonstrates that the inclusion of already weak intersite Coulomb
repulsion suppresses the Cooper pairing in $d+id-$wave channel at
the low densities, however it leads to realization of
$d+id-$pairing near the van Hove singularity
(Fig.~\ref{lambdas_ab}(b)). As a result, the $f-$wave pairing
takes place for the densities $1<n<1.21$. A further increase in
parameter $V$ leads to the growth of the pairing intensity in both
$f-$wave and $d+id-$wave channels (Fig.~\ref{lambdas_ab}(c)). The
leading SC-instability here corresponds to $f-$wave pairing for
the densities $1<n<1.23$ and to $d+id-$wave pairing near the van
Hove singularity. In the calculation of the dependencies of
$\lambda(n)$ in Fig.~\ref{lambdas_ab}(c), we used the parameters
close to those obtained from the ab initio calculation in
paper~\cite{Wehling11}.
\begin{figure}[h]
\begin{center}
\includegraphics[width=0.35\textwidth]{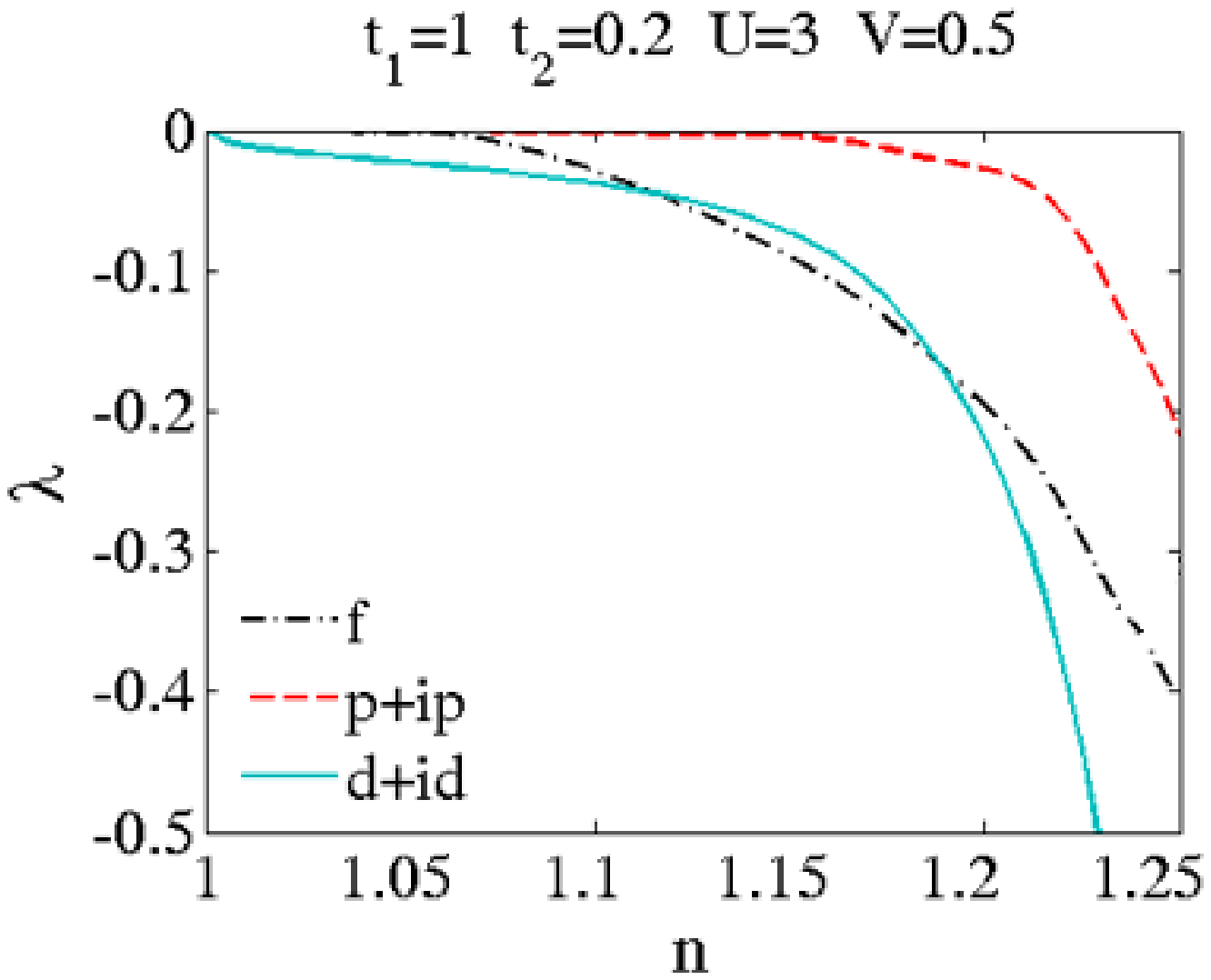}
\caption{\label{lambdas_ab_t2}Dependences of $\lambda$ on the
electron density $n$ with respect to the effective interaction of
electrons with the energies corresponding to both branches of the
graphene energy spectrum for $t_2=0.2|t_1|,\,U=3|t_1|$, and
$V=0.5|t_1|$.}
\end{center}
\end{figure}

The account for electron hoppings to the next to nearest-neighbor
carbon atoms ($t_2$) does not qualitatively affect the interplay
of the superconducting phases (Fig.~\ref{lambdas_ab}). This is
illustrated in Fig.~\ref{lambdas_ab_t2}, where we show the
dependences of $\lambda(n)$ for the parameters
$t_2=0.2|t_1|,\,U=3|t_1|$, and $V=0.5|t_1|$. Here again we have
$d+id-$wave pairing for $1<n<1.12$ and $1.18<n<1.25$, and $f-$wave
pairing for $1.12<n<1.18$. Such a behavior of the system is
explained by the fact that the inclusion of the hoppings $t_2>0$
or $t_2<0$ does not significantly modify the electron density of
states of graphene in the regions of the carrier concentration
between the Dirac point and both points of the van Hove
singularities $n_{vH}$ (Fig.~\ref{DOS}). However, it can be seen
in Fig.~\ref{lambdas_ab_t2} that the account for the hoppings
$t_2$ leads to an increase in the absolute values of the effective
interaction and, consequently, realization of the higher critical
temperatures of the transition to the superconducting phase in
graphene.

\section{Conclusion}
\label{Conclusion}

Considering a monolayer of an idealized graphene and neglecting
the van der Waals potential of a substrate and the role of the
nonmagnetic impurities, we demonstrated that the Kohn-Luttinger
superconducting pairing can be realized in the systems with the
linear dispersion law. The electronic structure of graphene was
described by the strong coupling Wannier representation on the
hexagonal lattice within the Shubin-Vonsowsky model, which takes
into account not only the intrasite, but also the intersite
Coulomb repulsion. We constructed the superconductive phase
diagram and demonstrated that the inclusion of the intersite
Coulomb repulsion significantly changes the regions occupied by
the superconducting phases with the $f-$, $d+id-$, and $p+ip-$wave
symmetries of the order parameter. On the other hand, the account
of the distant electron hoppings only weakly modifies the phase
diagram. At the same time, it leads to an increase in the absolute
values of the effective interaction and, consequently, to the
higher critical temperatures (up to $T\sim 10 K$) of the
superconducting transition in graphene.

It will be interesting to generalize our results on bilayer and
multilayer graphene structures. Note that rigorously speaking the
substantial difference between graphene and graphite manifests
itself only on the level of two layers and the rotation of their
elementary lattice cells with respect to each other. In bilayer
and multilayer graphene, it is very important and desirable
(see~\cite{Kagan91,Kagan11a,Kagan11b}) to take the interlayer
Hubbard repulsion $U_{12}$ into account and to construct the
superconducting phase diagram as a function of the relative
strength of the interlayer and onsite Hubbard repulsions $U_{12}/
U_1$ and relative electron densities in the layers $n_1/n_2$. It
will be also interesting to perform the calculations in the spirit
of experiments~\cite{Bozovic94} on high-$T_c$ superconducting
systems and to find a pronounced maximum in $T_c$ as a function of
the number of layers.

Acknowledgments

This work was supported by the Program of the Division of Physical
Sciences of the Russian Academy of Sciences (project II.3.1), and
the Russian Foundation for Basic Research (projects 14-02-00058
and 14-02-31237). One of authors (M.\,M.\,K.) acknowledges the
support of the Council of the President of the Russian Federation
(project MK-526.2013.2), and the Dynasty Foundation.






\begin{thebibliography}{6}
\bibitem{Lozovik08}
Yu.\,E. Lozovik, S.\,P. Merkulova, A.\,A. Sokolik, Phys. Usp.
\textbf{51} (2008) 727.

\bibitem{Castro09}
A.\,H. Castro Neto, F. Guinea, N.\,M.\,R. Peres, K.\,S. Novoselov,
A.\,K. Geim, Rev. Mod. Phys. \textbf{81} (2009) 109.

\bibitem{Heersche07}
H.\,B. Heersche, P. Jarillo-Herrero, J.\,B. Oostinga, L.\,M.\,K.
Vandersypen, A.\,F. Morpurgo, Nature (London) \textbf{446} (2007)
56.

\bibitem{Peres05}
N.\,M.\,R. Peres, F. Guinea, A.\,H. Castro Neto, Phys. Rev. B
\textbf{72} (2005) 174406.

\bibitem{Marino06}
E.\,C. Marino, L.\,H.\,C.\,M. Nunes, Nucl. Phys. B \textbf{741}
(2006) 404.

\bibitem{Gonzalez01}
J. Gonz\'{a}lez, F. Guinea, M.\,A.\,H. Vozmediano, Phys. Rev. B
\textbf{63} (2001) 134421.

\bibitem{Uchoa07}
B. Uchoa, A.\,H. Castro Neto, Phys. Rev. Lett. \textbf{98} (2007)
146801.

\bibitem{Black07}
A.\,M. Black-Schaffer, S. Doniach, Phys. Rev. B \textbf{75} (2007)
134512.

\bibitem{Honerkamp08}
C. Honerkamp, Phys. Rev. Lett. \textbf{100} (2008) 146404.

\bibitem{Kiesel12}
M.\,L. Kiesel, C. Platt, W. Hanke, D.\,A. Abanin, R. Thomale,
Phys. Rev. B \textbf{86} (2012) 020507(R).

\bibitem{Gonzalez08}
J. Gonz\'{a}lez, Phys. Rev. B \textbf{78} (2008) 205431.

\bibitem{Markiewicz97}
R.\,S. Markiewicz, J. Phys. Chem. Solids \textbf{58} (1997) 1179.

\bibitem{Kohn65}
W. Kohn, J.\,M. Luttinger, Phys. Rev. Lett. \textbf{15} (1965)
524.

\bibitem{Valenzuela08}
B. Valenzuela, M.\,A.\,H. Vozmediano, New J. Phys. \textbf{10}
(2008) 113009.

\bibitem{Nandkishore12}
R. Nandkishore, L.\,S. Levitov, A.\,V. Chubukov, Nature Phys.
\textbf{8} (2012) 158.

\bibitem{Wehling11}
T.\,O. Wehling, E. \c{S}a\c{s}{\i}o\u{g}lu, C. Friedrich, A.\,I.
Lichtenstein, M.\,I. Katsnelson, S. Blugel, Phys. Rev. Lett.
\textbf{106} (2011) 236805.

\bibitem{Shubin34}
S. Shubin, S. Vonsowsky, Proc. Roy. Soc. A {\bf 145} (1934) 159.

\bibitem{Wallace47}
P.\,R. Wallace, Phys. Rev. \textbf{71} (1947) 622.

\bibitem{Gor'kov61}
L.\,P. Gor'kov, T.\,K. Melik-Barkhudarov, Sov. Phys. JETP {\bf 13}
(1961) 1018.

\bibitem{Scalapino86}
D.\,J. Scalapino, E. Loh, Jr., J.\,E. Hirsch, Phys. Rev. B
\textbf{34} (1986) 8190.

\bibitem{Baranov92}
M.\,A. Baranov, A.\,V. Chubukov, M.\,Yu. Kagan, Int. J. Mod. Phys.
B \textbf{6} (1992) 2471.

\bibitem{Hlubina99}
R. Hlubina, Phys. Rev. B \textbf{59} (1999) 9600.

\bibitem{Raghu10}
S. Raghu, S.\,A. Kivelson, D.\,J. Scalapino, Phys. Rev. B {\bf 81}
(2010) 224505.

\bibitem{Kagan13a}
M.\,Yu. Kagan, V.\,V. Val'kov, V.\,A. Mitskan, M.\,M. Korovushkin,
JETP Lett. {\bf 97} (2013) 226.

\bibitem{Kagan13b}
M.\,Yu. Kagan, V.\,V. Val'kov, V.\,A. Mitskan, M.\,M. Korovushkin,
JETP {\bf 117} (2013) 728.

\bibitem{Landau77}
L.\,D. Landau, E.\,M. Lifshitz, Quantum Mechanics:
Non-Relativistic Theory, Pergamon, Oxford, 1977.

\bibitem{Dzyaloshinskii88a}
I. E. Dzyaloshinskii, V. M. Yakovenko, Sov. Phys. JETP \textbf{67}
(1988) 844.

\bibitem{Dzyaloshinskii88b}
I.\,E. Dzyaloshinskii, I.\,M. Krichever, Ya.\,Khronek, Sov. Phys.
JETP \textbf{67} (1988) 1492.

\bibitem{Kagan91}
M.\,Yu. Kagan, Phys. Lett. A \textbf{152} (1991) 303.

\bibitem{Kagan11a}
M.\,Yu. Kagan, V.\,V. Val'kov, JETP \textbf{113} (2011) 156.

\bibitem{Kagan11b}
M.\,Yu. Kagan, V.\,V. Val'kov, Sov. Phys. Low Temp. \textbf{37}
(2011) 84.

\bibitem{Bozovic94}
I. Bozovic, J.\,N. Eckstein, G.\,F. Virshup, Physica C
\textbf{235-240} (1994) 178.


\end{thebibliography}







\end{document}